\documentclass[a4paper]{article}

\usepackage{INTERSPEECH2020}
\usepackage{ctable,multirow,graphicx,caption,subcaption,dblfloatfix,relsize,hyperref}

\title{SkipConvNet: Skip Convolutional Neural Network for Speech Dereverberation using Optimally Smoothed Spectral Mapping}
\name{Vinay Kothapally$^1$, Wei Xia$^1$, Shahram Ghorbani$^1$, John H.L. Hansen$^1$ \\
Wei Xue$^2$, Jing Huang$^2$}
\address{ \vspace{-0.5em}
  $^{1}$Center for Robust Speech Systems (CRSS), The University of Texas at Dallas, TX, USA \\
  $^{2}$JD AI Research, JD.com, USA}
\email{\{vinay.kothapally, wei.xia, shahram.ghorbani,john.hansen\}@utdallas.edu \\ \{xuewei27, jing.huang\}@jd.com}

\begin{document}
\vspace{-0.5em}
\maketitle

%=============================== ABSTRACT =============================%
\begin{abstract}
\vspace{-0.5em}
The reliability of using fully convolutional networks (FCNs) has been successfully demonstrated by recent studies in many speech applications. One of the most popular variants of these FCNs is the `U-Net', which is an encoder-decoder network with skip connections. In this study, we propose `SkipConvNet' where we replace each skip connection with multiple convolutional modules to provide decoder with intuitive feature maps rather than encoder's output to improve the learning capacity of the network. We also propose the use of optimal smoothing of power spectral density (PSD) as a pre-processing step, which helps to further enhance the efficiency of the network. To evaluate our proposed system, we use the REVERB challenge corpus to assess the performance of various enhancement approaches under the same conditions. We focus solely on monitoring improvements in speech quality and their contribution to improving the efficiency of back-end speech systems, such as speech recognition and speaker verification, trained on only clean speech. Experimental findings show that the proposed system consistently outperforms other approaches.

% This was used along with the generative adversarial network (GAN) in enhancing speech deteriorated by the presence of reverberation.

\end{abstract}

%=============================== INTRODUCTION =============================%
\section{Introduction}

Recent years have seen an exponential rise in the need for effective distant speech systems to improve the experience of human-machine interactions. These systems find their applications in many consumer devices today as personal assistance. Speech captured by these devices in confined spaces such as conference rooms, lobby, cafeteria, etc. faces two major challenges: (a) reverberation: self-distortion due to reflections which greatly reduce the intelligibility of the speech, and (b) background-noise: speech from multiple overlapping speakers, music or other acoustical sounds picked up from the environment. These two challenges are well-known and have been dealt with various signal processing and deep neural network (DNN) based speech enhancement strategies.

In earlier days, statistical signal enhancement methods played a crucial part in all speech processing pipelines \cite{dereverb1,dereverb2,dereverb3}.  However, over the past few years, many DNN based approaches for speech enhancement showed promising results in enhancing a reverberant speech. While DNN strategies for time-domain and frequency-domain processing \cite{KeTan,Ashutosh} were developed, most approaches prefer to operate on short-time fourier transform (STFT) of reverberant speech, to enhance the log-power spectrum (LPS) and reuse the unaltered noisy phase signal to restore a clean time-domain signal. As reverberation  has  its  effects  spread  over  time  and  frequency,  sequence-to-sequence learning strategies like recurrent neural networks (RNNs) and long short-term  memory (LSTM) \cite{LSTM} have been explored to capture and leverage the temporal correlations for speech  dereverberation. Besides the extreme capabilities of these networks to capture temporal correlations in speech, they fail to capture the spectral structure of formants encoded in the short-time fourier transform (STFT). Therefore, researchers have moved to convolutional neural networks (CNNs) which learn the dependencies from a group of neighboring time-frequency pixels. In a conventional CNN, the spectral structure learned using 2-D convolutions is compromised due to the presence of fully connected layers. For this reason, researchers used fully convolutional networks (FCNs) which substitute the fully connected layers with 1x1 convolutions in order to prevent the loss of spectral structure information. In past couple of years, many FCN architectures like U-Net, ResNet, DenseNet etc. were adopted from computer vision for various speech applications \cite{UNet,UNet+,ResNet,DenseNet,InceptionNet}. Since these networks have shown significant success, exploration of various network architectures that further improve the system performance in speech related tasks has been a part of the research. In this study, we propose modifications to one such FCN architecture, U-Net, specifically designed for speech dereveberation task. We also show that using pre-processed LPS for training such networks improves the efficiency significantly.

The rest of this paper is organized as follows. Section 2 briefly introduces to the problem statement.  Section-3, provides insights on the optimal smoothing proposed to be used as a pre-processing step.  Section-4, describes the proposed `SkipConvNet' for single-channel speech dereverberation. Details on the experimental setup and results are presented in Section-5. Finally, we conclude our work in Section 6.

\renewcommand{\thefigure}{2}
\begin{figure*}[b!]
  \includegraphics[width=\textwidth]{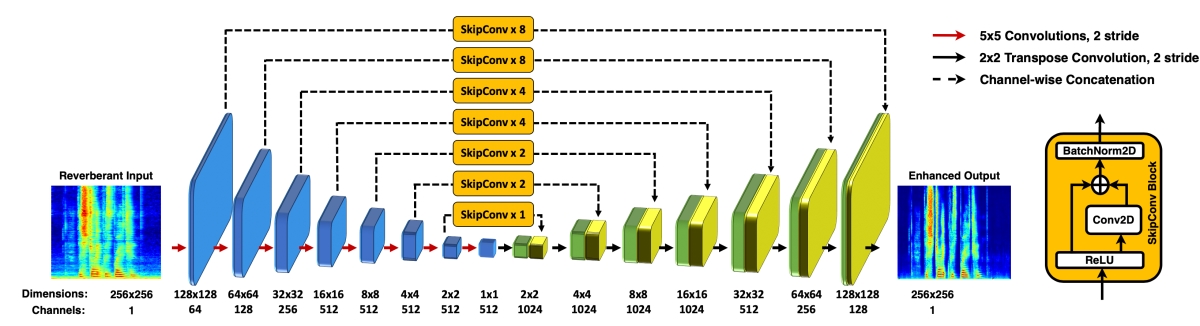}
  \vspace{-2em}
  \caption{SkipConvNet: UNet with convolution modules in skipped connections trained on optimally smoothed PSDs }
  \label{fig:Network}
\end{figure*}

%=============================== PROBLEM STATEMENT =============================%
\section{Problem Formulation}

For a given a room impulse response (RIR), a reverberant speech signal received by an omni-directional microphone can be modeled as:
\vspace{-0.8em}
\begin{equation}
    \label{conv}
    x(t) = \sum_{t=0}^{L}{s(t)*h(L-t)} + n(t)
    \vspace{-0.3em}
\end{equation}
\begin{equation}
    \centering
    \label{conv_fft}
    X(t,f) = S(t,f)H(t,f) + N(t,f)
    \vspace{0.4em}
\end{equation}

\noindent where, $x(t)$ is the signal as observed by a distant microphone, $s(t)$ is the clean speech signal from the source,  $h(t)$ is the room impulse response (RIR), and $n(t)$ is background additive noise. The relation in frequency domain can be represented as Eq-(\ref{conv_fft}), where $X(t,f), S(t,f)$, $H(t,f)$ and $N(t,f)$ represent the STFT of observed reverberated speech, clean speech, RIR and background noise respectively. In this study, the noise level is considerably lower than the target speaker so that the analysis concentrates solely on evaluating and suggesting strategies to reverse the impacts of reverberation. The early and late reflections in a room impulse response (RIR) creates a smearing effect both in time and frequency of a speech spectrogram, see Fig-\ref{fig:Optimal_smoothing}(a). The emphasis of this study is on learning a non-linear function using FCNs that maps the LPS of a noisy \& reverberant speech $X(t,f)$ to a corresponding clean speech $S(t,f)$. Later, the estimated enhanced LPS from the network is combined with the unaltered noisy phase response to reconstruct the enhanced speech.

%=============================== Preprocessing =============================%

\section{Optimal Smoothing based Pre-processing}
Estimation of the effects of late reflections and/or background noise statistics plays an important role in the design of a robust speech enhancement system. We use a minimum statistics based approach \cite{minStats} originally used to estimate the power spectral density (PSD) of noise, to also estimate the same for late reflections given noisy \& reverberant speech. The approach is based on the assumption that energy across all frequencies will theoretically tend to be zero during silent periods in speech, unless affected by stationary-noise or reverberation or both. Thus, tracking the minimum energy in each frequency bin over time for a smoothed reverberant PSD can produce a robust estimate of the PSD of noise/reverberation. Although this approach is focused on estimating PSD of noise/reverberation, we discard it and use only the smoothed speech PSD for training our proposed system.

\vspace{-0.8em}
\begin{equation}
    \centering
    \label{psd_smoothing}
    \resizebox{.9\hsize}{!}{$P(t,f) = \alpha_{opt}(t,f)P(t-1,f) + (1-\alpha_{opt}(t,f)|X(t,f)|^2,$}
\end{equation}
\vspace{-2em}

\begin{equation}
    \centering
    \label{optimal_alpha}
    \resizebox{.68\hsize}{!}{$ \alpha_{opt}(t,f) = \frac{1}{1+[P(t-1,f)/\sigma^2_{n}(t,f)-1]^2}.$}
\end{equation}
\vspace{0.1em}

In general, a fixed smoothing parameter `$\alpha$' is used in speech applications to obtain a smoothed PSD, $P(t,f)$. It is known that having a fixed value often comes with trade-off issues in estimation. A sliding window smoothing mechanism is robust for a higher value of `$\alpha$' but blurs the speech activity and silence boundaries. On the contrary, abrupt changes in speech activity can be recorded with a lower value of `$\alpha$' at the cost of a less reliable estimate of PSD. To address this issue, a time-varying and frequency-dependent smoothing parameter is used to get an accurate estimate of speech PSD as shown in Eq-(\ref{psd_smoothing}). The optimal smoothing parameter is computed using Eq-(\ref{optimal_alpha}), where $\sigma^2_{n}(t,f)$ is the variance of noise. We suggest referring \cite{minStats} for a detailed derivation of this optimal smoothing parameter. In addition, the smooth PSD values below `-80 dB' for all training samples are clipped to have a constant dynamic range, see Fig-\ref{fig:Optimal_smoothing}.

\renewcommand{\thefigure}{1}
\begin{figure}[h!]
    \centering
    \includegraphics[width=\linewidth]{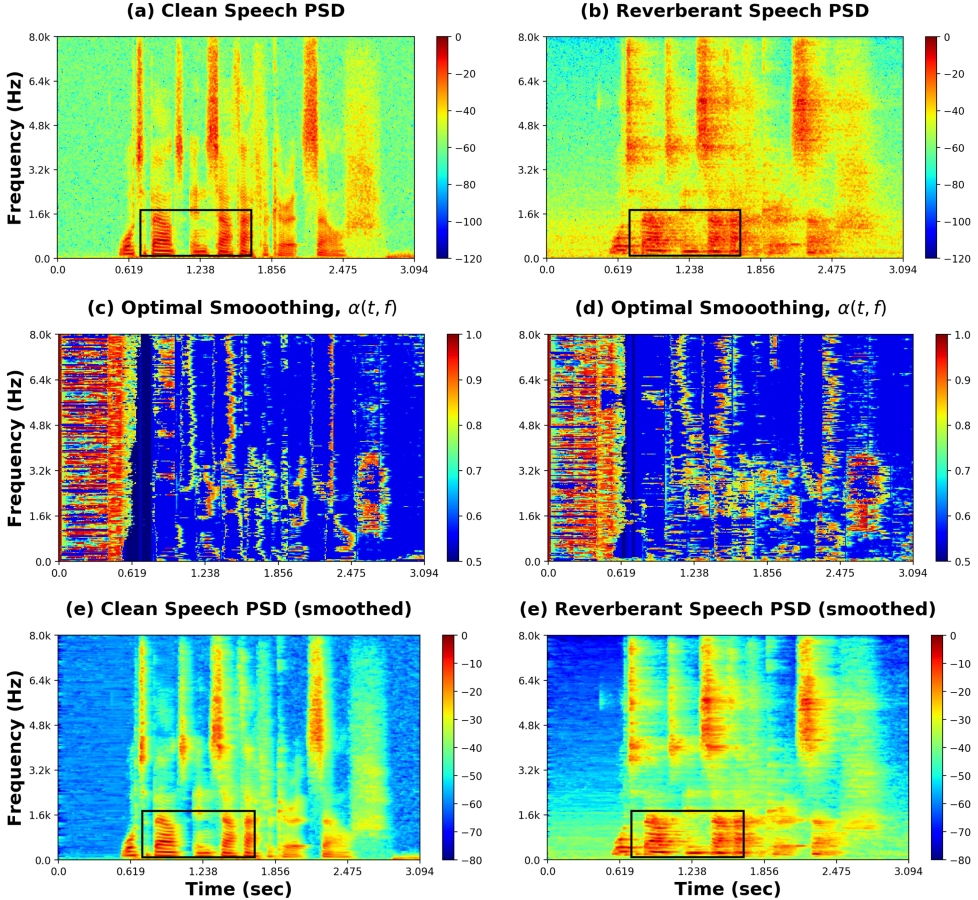}
    \vspace{-1em}
    \caption{Optimal Smoothing as Preprocessing}
    \label{fig:Optimal_smoothing}
\end{figure}

The original and smoothed versions of PSD for a reverberant and a corresponding clean speech utterances are shown in Fig-\ref{fig:Optimal_smoothing}. The optimal smoothing parameter adapts itself accordingly for active and silent (late reflections in our case) regions of speech. It is clear from the highlighted regions in Fig-\ref{fig:Optimal_smoothing} that optimal smoothing helps retrieve the lost formant structure in reverberant speech. Thus, we propose to use this smoothing strategy to pre-process the reverberant speech before being fed to the proposed system for training purposes.

\renewcommand{\thefigure}{3}
\begin{figure*}[!b]
  \centering
  \includegraphics[width=\textwidth]{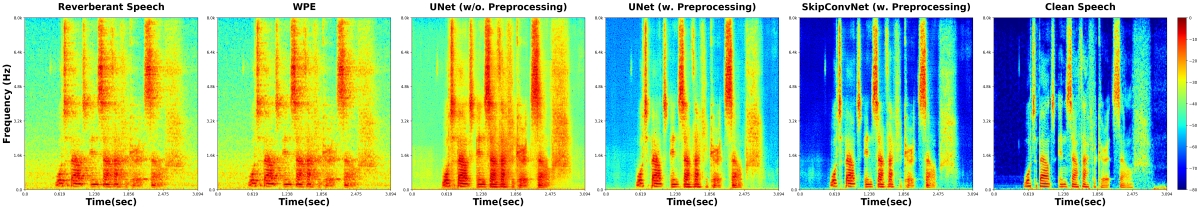}
  \vspace{-0.5em}
  \caption{Reverberant, Enhanced and Clean Speech Spectrograms}
  \label{fig:Enhanced Images}
\end{figure*}
\vspace{-0.2em}

\section{Skip Convolutional Neural Network}
In this section, we start with a formal description of standard U-Net and then explain the modifications we propose to make it a `SkipConvNet'. U-Net is an encoder-decoder based image-2-image translation network. The encoder stage in the architecture extracts spectral and temporal features from the LPS of the input reverberant speech, and the decoder constructs an enhanced LPS from the encoded features. The encoder and decoder networks constitute of multiple layers of convolutions followed by down-sampling and up-sampling, respectively. With increasing numbers of layers in the network, each neuron's reception field increases. Alternatively, the dimension of the processed encoded features decreases. Similar to \cite{Ernst}, we use enough layers in the architecture such that the encoder downsamples a given input to a single pixel. This ensures that the decoder uses all spectral and temporal features learned by the encoder from the input to construct an enhanced output. The skip connections play the most crucial role in the U-Net architecture. A skip connection is a link between the encoder and decoder used to share learned features, represented by dotted lines in Fig-\ref{fig:Network}. A skip connection in each layer concatenates the output from encoder before down-sampling with the decoder in a corresponding layer, with the assumption that both input and the output PSD have a similar structure. These concatenated features from the encoder and the previous layers of the decoder not only helps to preserve the information from being lost during down-sampling in the encoder, but also guides the decoder towards the reconstruction of the enhanced output.

Although the skip connections have proven to be efficient in building a robust system, a recent study by \cite{MultiResUNet} analyzes a probable semantic gap in features exchanged between encoder and decoder. For instance, the first layer of the encoder extracts low-level local spectral and temporal features. These features are concatenated with the final layer of the decoder which receives highly processed features from its previous layers. Merging these two incompatible sets of features might limit the learning abilities of the FCNs. Addition of a few convolutional layers within each skip connection can compensate for the incompatibilities by transforming the features from the encoder to be more intuitive to the decoder. We believe that, with the minimized differences within the features at each layer of the decoder, the learning ability of FCN's can potentially be maximized.

Unlike \cite{MultiResUNet}, which uses convolutions with varying kernels in parallel, we use standard convolutions followed by normalization and non-linear activation for the encoder and the decoder respectively. However, we use multiple `\textit{skipconv}' blocks in series for each skip connection in the architecture. A \textit{skipconv} block constitutes of a non-linear activation followed by a 5x5 convolution with a residual connection and does not alter the dimensions of the features set. The features are then normalized before being shared with either another \textit{skipconv} block or the decoder. The number of \textit{skipconv} blocks used in a particular skip connection is made to vary inversely with the depth of the layer in the encoder it is associated with, see Fig-\ref{fig:Network}. For instance, skip connection associated with the final layer of the encoder has only one \textit{skipconv} block whereas the first layer in the encoder has a total of eight \textit{skipconv} blocks. This is based on the assumption that the deeper layers of the network deal with high-level information and require minimal transformation to reduce the incompatibilities within feature sets at the decoder.

\section{Experimental Results}

All experiments were run on the REVERB challenge corpus \cite{ReverbChallenge1,ReverbChallenge2}. The outcomes of SkipConvNet are compared with the outcomes of a standard U-Net \cite{Ernst} trained on LPS of reverberant and clean speech utterances with and without the proposed pre-processing and a widely used statistic dereverberation algorithm, Weighted prediction error WPE \cite{WPE1,WPE2}. We also test the enhanced speech utterances with back-end speech automatic speech recognition (ASR) and speaker verification (SV) models trained using only clean speech.

\vspace{-0.5em}
\subsection{Dataset}
The Reverb Challenge corpus is a collection of simulated and real recordings of speech sampled at 16kHz from different rooms with varying levels of reverberation and a background noise at 20dB SNR. The simulated data is generated by convolving clean speech utterances from WSJCAM0 \cite{wsjcam0} and room impulse responses (RIRs) collected from three different rooms (small, medium and large sizes) and two different microphone placements (near, far) using a single microphone, 2-channel and an 8-channel microphone array. The multi-channel real recordings were drawn from MC-WSJ-AV corpus \cite{mcwsjav}. The corpus is divided into \textit{train}, \textit{dev} and \textit{eval} sets. The \textit{train} set consists of 7,861 simulated reverberant utterances which are used to train the systems being compared in this study. However, the \textit{dev} and \textit{eval} set contain both simulated and real recordings.

\vspace{-0.5em}
\subsection{Pre-Processing and Network architecture}
We compute STFT with a frame length of 512 samples and an overlap of 384 samples for a given speech utterance. We then compute LPS of a speech signal from an optimally smoothed PSD of the speech using Eq-\ref{psd_smoothing} \& \ref{optimal_alpha}. We only consider the lower half of the , since the STFT is symmetric.
Late, the LPS of each utterance is divided into batches with 256 consecutive frames to form spectral images of size 256x256. We re-use the U-Net architecture proposed in \cite{Ernst} as our baseline. Fig-\ref{fig:Network} gives an overview of the proposed `SkipConvNet' architecture with convolutional module `SkipConv block' added to the skip connections. All convolutions in the encoder and skipconv blocks use a kernel size of 5x5 with a stride of 2. Similarly, all convolutions in the decoder use transposed convolution with a kernel size and stride of 2. We train our network using a total of 62,888 spectral images (corresponding to 7,861 utterances) from \textit{train} set of the corpus with Adam optimizer. The netowrk is tarined to minimize the mean square error (MSE) between the network prediction and the LPS of corresponding clean utterance. We use a batch-size of 8 and train the network for 10 epochs. Finally, estimated LPS from the network is combined with the unaltered noisy phase to reconstruct the enhanced speech. We report the improvements seen on the \textit{eval} set of the corpus.

\begin{table*}[h!]
  \caption{Improvements in Speech Quality Measures for \textit{SimData} and \textit{RealData}}
  \vspace{-0.8em}
  \label{tab:Speech_Quality}
  \centering
  \resizebox{0.95\textwidth}{!}{
  \setlength{\tabcolsep}{8pt}
  \begin{tabular}{|l|c|c|c|c|c|c|c|c|c|c|c|c||c|}
    \hline
    ~ & \multicolumn{12}{c||}{\textbf{Simulated}} & \textbf{Real} \\[0.5ex] \cline{2-14}
    ~ & \multicolumn{3}{c|}{\textbf{CD}} & \multicolumn{3}{c|}{\textbf{LLR}} & \multicolumn{3}{c|}{\textbf{FWSegSNR}} & \multicolumn{3}{c||}{\textbf{SRMR}} & \multirow{ 2}{*}{\textbf{SRMR}}\\[0.5ex]
    \textbf{Room} & \multicolumn{1}{c}{\#1} & \multicolumn{1}{c}{\#2} & \multicolumn{1}{c|}{\#3} & \multicolumn{1}{c}{\#1} & \multicolumn{1}{c}{\#2} & \multicolumn{1}{c|}{\#3} & \multicolumn{1}{c}{\#1} & \multicolumn{1}{c}{\#2} & \multicolumn{1}{c|}{\#3} & \multicolumn{1}{c}{\#1} & \multicolumn{1}{c}{\#2} & \multicolumn{1}{c||}{\#3} & ~\\
    \hline
    \multicolumn{14}{|c|}{Far Microphone}\\[0.5ex]
    \hline
    Reverb & 2.65 & 5.08 & 4.82 & 0.38 & 0.77 & 0.85 & 6.75 & 0.53 & 0.14 & 4.63 & 2.94 & 2.76 & 3.51\\
    WPE & 2.42 & 5.04 & 4.76 & 0.35 & 0.79 & 0.83 & 7.34 & 0.80 & 0.34 & 4.87 & 3.15 & 2.99 & 3.85\\
    UNet & 2.18 & 3.65 & 3.42 & 0.26 & 0.60 & 0.56 & 7.19 & 2.16 & 2.13 & 4.54 & 3.49 & 3.30 & 4.40\\
    UNet+Pre-Processing & 2.28 & 3.27 & 3.07 & 0.25 & 0.49 & 0.48 & 10.47 & 7.73 & 6.93 & 4.84 & 4.42 & 4.08 & 5.51\\
    SkipConvNet & \textbf{2.12} & \textbf{3.06} & \textbf{2.82} & \textbf{0.22} & \textbf{0.46} & \textbf{0.46} & \textbf{11.80} & \textbf{8.88} & \textbf{8.16} & \textbf{5.10} & \textbf{4.76} & \textbf{4.25} & \textbf{6.87}\\
    % \specialrule{.05em}{.05em}{.05em}
    \hline
    \multicolumn{14}{|c|}{Near Microphone}\\[0.5ex]
    \hline
    % \specialrule{.05em}{.05em}{.05em}
    Reverb & 1.96 & 4.58 & 4.20 & 0.34 & 0.51 & 0.65 & 8.10 & 3.07 & 2.32 & 4.37 & 3.67 & 3.66 & 4.05\\
    WPE & 1.82 & 4.53 & 4.12 & 0.33 & 0.52 & 0.62 & 8.66 & 3.44 & 2.69 & 4.51 & 3.89 & 3.92 & 4.42\\
    UNet & 2.14 & 3.06 & 2.79 & 0.28 & 0.39 & 0.38 & 7.31 & 4.41 & 5.24 & 4.34 & 3.91 & 4.06 & 4.68\\
    UNet+Pre-Processing & 2.05 & 2.82 & 2.71 & 0.22 & 0.35 & 0.36 & 11.68 & 9.62 & 8.87 & 4.67 & 4.50 & 4.40 & 5.87\\
    SkipConvNet & \textbf{1.86} & \textbf{2.57} & \textbf{2.45} & \textbf{0.19} & \textbf{0.30} & \textbf{0.35} & \textbf{13.07} & \textbf{10.96} & \textbf{10.22} & \textbf{4.99} & \textbf{4.75} & \textbf{4.56} & \textbf{7.27}\\
    \hline
    \end{tabular}
    }
\end{table*}

\begin{table*}[h!]
  \caption{SV and ASR performance on simulated and real recordings for models trained on clean speech}
  \label{tab:Backend_Systems}
  \vspace{-0.8em}
  \centering
  \resizebox{0.85\textwidth}{!}{
  \setlength{\tabcolsep}{8pt}
  \begin{tabular}{|l|c|c|c|c|c|c||c|c|c|c|c|c|}
    \hline
    ~ & \multicolumn{6}{c||}{\underline{\textbf{Speaker Verification}}} & \multicolumn{6}{c|}{\underline{\textbf{Speech Recognition}}} \\[0.5ex] %\cline{2-13}
    ~ & \multicolumn{6}{c||}{EER (\%) X-vector PLDA-trained on Clean Speech } & \multicolumn{6}{c|}{WER (\%) Acoustic Model-trained on Clean Speech} \\[0.5ex] \cline{2-13}
    ~ & \multicolumn{3}{c|}{\textbf{SimData}} &\multicolumn{3}{c||}{\textbf{RealData}} & \multicolumn{3}{c|}{\textbf{SimData}} &\multicolumn{3}{c|}{\textbf{RealData}} \\[0.5ex]
    \textbf{Method} & \multicolumn{1}{c}{1ch} & \multicolumn{1}{c}{2ch} & \multicolumn{1}{c|}{8ch}  & \multicolumn{1}{c}{1ch} & \multicolumn{1}{c}{2ch} & \multicolumn{1}{c||}{8ch} & \multicolumn{1}{c}{1ch} & \multicolumn{1}{c}{2ch} & \multicolumn{1}{c|}{8ch}  & \multicolumn{1}{c}{1ch} & \multicolumn{1}{c}{2ch} & \multicolumn{1}{c|}{8ch} \\ \hline
    Reverb & 8.21 & - & - & 6.14 & - & -  &  34.92 & - & - & 93.52 & - & - \\
    WPE & 8.51 & 5.88 & 1.72 & 6.14 & 5.26 & 4.09 &  30.32 & 18.76 & \textbf{5.79} & 90.19 & 78.45 & 53.78 \\
    UNet+Pre-Processing & 2.63 & 2.01 & 1.86 & 5.85 & 5.56 & 4.39 & 10.67 & 9.26 & 7.19 &   48.33 & 49.12 & 46.79 \\
    SkipConvNet & \textbf{2.20} & \textbf{2.01} & \textbf{1.48} & \textbf{5.26} & \textbf{3.80} & \textbf{3.51}  & \textbf{8.99} & \textbf{7.54} & 5.81 & \textbf{35.73} & \textbf{34.22} & \textbf{30.99} \\
    \hline
    \end{tabular}
    }
\end{table*}

\vspace{-0.5em}

\subsection{Results}
We begin our presentation of experimental results with a second look at the optimal smoothed PSD's from Fig-\ref{fig:Optimal_smoothing}. From Fig-\ref{fig:Optimal_smoothing}(a),(c) \& (e), we see that optimal smoothing helps in preserving the formant structure during the speech frames by having a low smoothing parameter while assigning the regions with reverberant contents with a higher smoothing parameter.

We then measure the relative enhancement achieved by each system using several speech quality measures, as shown in Table-\ref{tab:Speech_Quality}. A FCN based U-Net and the proposed `SkipConvNet' performed consistently better compared to the widely used statistical dereverberation algorithm, WPE. However, we observed a 39.19\% relative improvement in the performance of the baseline U-Net by solely introducing the proposed pre-processing. This shows that the proposed pre-processing helps all FCN networks and is not biased to the proposed `SkipConvNet'. However, the proposed `SkipConvNet' consistently performed the best compared to U-Net and the U-Net trained on pre-processed inputs with an average of 54.45\% and 10.40\% relative improvements over all quality metrics respectively. Consistent improvements in `SRMR' and `FWSegSNR' in addition to `CD' ensure the reduction of reverberation and background noise in enhanced speech utterances without any processing artifacts/distortions.

Finally, we test the improvements in back-end automatic speech recognition (ASR) and speaker verification systems (SV) achieved with proposed system for single and multi-channel streams, see Table-\ref{tab:Backend_Systems}. For multi-channel streams of data, individual channels are enhanced with different dereveberation techniques discussed in the study and then spatially combined using BeamformIT \cite{Beamform1,Beamform2} beamforming strategy. Since the proposed pre-processing enhanced the performance of traditional U-Net, we compare the proposed system's achievements with only WPE and U-Net trained on pre-processed spectral images. For an ASR system, we use TDNN based acoustic model \cite{KaldiASR} trained on single channel clean speech of the REVERB CHALLENGE corpus. Similarly, for a speaker verification system, we train a X-vector model \cite{xvectors} on Voxceleb 1 \& 2 corpus \cite{VoxCeleb1,VoxCeleb2} and a PLDA backend on the in-domain single-channel clean speech of the corpus. Three utterances from each speaker from both simulated and real recordings of \textit{eval} are considered in the enrollment set and the rest in the evaluation set.
We see a relative improvement of 35.03\% and 16.42\% in speaker verification performance using X-vectors averaged over simulated and real recordings compared to WPE and U-Net trained on pre-processed spectral images. Similarly, we see a 48.15\% \& 23.94\% relative improvements in the performance of an automated speech recognition (ASR) system averaged over simulated and real recordings compared to WPE and U-Net trained on pre-processed spectral images. The interested reader may check out some audio samples at: \url{https://vkothapally.github.io/SkipConv/}

\vspace{-0.5em}
\section{Conclusions}

In this study, we presented `SkipConvNet' an encoder-decoder based FCN with convolutional modules introduced in skip connections which enhanced the learning ability to map reverberant speech to its corresponding clean speech. We have proposed the use of optimal smoothing of PSD as a preprocessing step for training the network which has shown considerable improvements in the network's performance.
With the proposed modifications, we achieved significant improvements on speech quality for both real and simulated data from REVERB CHALLENGE corpus in comparison with traditional U-Net and also widely-used WPE dereverberation algorithm.
We have also shown that the proposed system also improves the performance of single-channel and multi-channel back-end speech systems like speech recognition and speaker verification. To summarize, the addition of convolutions in skip connections reduces the incompatibilities within the feature sets received at each layer of the decoder and boosts the learning capabilities of the network. We believe that this work can potentially be extended to a number of complex-FCN architectures that have recently been researched for speech enhancement.

\newpage
\bibliographystyle{IEEEtran}

\bibliography{mybib}

\end{document}